# A hybrid polymer/ceramic/semiconductor fabrication platform for high-sensitivity fluid-compatible MEMS devices with sealed integrated electronics

Nahid Hosseini[1†], Matthias Neuenschwander[1†], Jonathan D. Adams[1], Santiago H. Andany[1], Oliver Peric[1], Marcel Winhold[2], Maria Carmen Giordano[3], Vinayak Shantaram Bhat[3], Dirk Grundler[3], and Georg E. Fantner[1*]

[1]Laboratory for Bio- and Nano-Instrumentation, École Polytechnique Fédérale de Lausanne (EPFL), Switzerland. [2]GETec Microscopy GmbH, Vienna, Austria. [3] Laboratory of Nanoscale Magnetic Materials and Magnonics, École Polytechnique Fédérale de Lausanne (EPFL)

† These authors contributed equally to this work

* Corresponding author

## Abstract

Active microelectromechanical systems can couple the nanomechanical domain with the electronic domain by integrating electronic sensing and actuation mechanisms into the micromechanical device. This enables very fast and sensitive measurements of force, acceleration, or the presence of biological analytes. In particular, strain sensors integrated onto MEMS cantilevers are widely used to transduce an applied force to an electrically measurable signal in applications like atomic force microscopy, mass sensing, or molecular detection. However, the high Young's moduli of traditional cantilever materials (silicon or silicon nitride) limit the thickness of the devices, and therefore the deflection sensitivity that can be obtained for a specific spring constant. Using softer materials such as polymers as the structural material of the MEMS device would overcome this problem. However, these materials are incompatible with high-temperature fabrication processes often required to fabricate high quality electronic strain sensors. We introduce a pioneering solution that seamlessly integrates the benefits of polymer MEMS technology with the remarkable sensitivity of strain sensors, even under high-temperature deposition conditions. Cantilevers made using this technology are inherently fluid compatible and have shown up to 6 times lower force noise than their conventional counterparts. We demonstrate the benefits and versatility of this polymer/ceramic/semiconductor multi-layer fabrication approach with the examples of self-sensing AFM cantilevers, and membrane surface stress sensors for biomolecule detection.

**Introduction**

Microelectromechanical systems (MEMS) have become highly successful in sensing applications in nanoscience. For example, AFM cantilevers have a very simple geometry, and yet are capable of probing samples in the nanoscale regime with remarkable resolution. Traditionally, the deflection of AFM cantilevers is detected using the optical beam deflection (OBD) method[1] where a laser beam is reflected from the back of the cantilever and centred on a quadrant photodiode (Figure 1a). Soon after the invention of the AFM, cantilevers with integrated sensing elements were proposed for various applications[2–8]. Such self-sensing cantilevers are usually made of traditional MEMS materials (silicon or silicon nitride[9,10]) and feature a piezoresistive strain sensor near their fixed end (Figure 1b). However, they have not found widespread use in AFM, because of their lower force sensitivity and signal to noise ratio compared to optically detected cantilevers. This results from the fundamentally different quantities being measured by the two methods. Self-sensing cantilevers measure the strain in the base of the cantilever resulting from an applied force. In contrast, optically detected cantilevers measure a change in the angle of the cantilever at its free end. In both cases, the achievable force sensitivity (FS) depends on the deflection sensitivity (DS) of the readout method and the cantilever spring constant (k), FS=DS/k. For a given deflection at the free end of the cantilever, the deflection angle does not depend on the thickness of the cantilever (supplementary material S1). For optical deflection detection, the deflection sensitivity is therefore independent of the cantilever thickness (see figure 1c and supplementary material S1). The deflection sensitivity of strain-based, self-sensing MEMS cantilevers, however, increases with the thickness of the cantilever (figure 1c and supplementary material S2). As such, a thicker cantilever exhibits a higher deflection sensitivity[11]. However, the spring constant increases strongly with the thickness of the cantilever. To obtain a higher force sensitivity, the thickness of the cantilever should be increased, without increasing the spring constant. This could be done by using materials with a lower Young's modulus such as polymers.

In recent years, polymer MEMS have shown to be attractive for numerous applications[12–15]. The low Young's moduli of polymers (e.g., the Young's modulus of SU-8 is ca. 60 times lower than silicon nitride) allows for thicker cantilevers while maintaining a low spring constant (see figure 1d). This softness makes polymer cantilevers particularly well-suited for self-sensing detection.

Johanson et al. reported the successful integration of metal strain gauges on SU-8 cantilevers[12], and the use of conducting polymers as strain sensors has been explored[16]. However, the achievable gauge factors for such sensors (for example, 2 for gold and -5 for polyaniline) are far below those of high-temperature-deposited semiconductor strain sensors (for example, 100 for doped single crystal silicon). The advantage of increased cantilever thickness is therefore overshadowed by lower strain sensor performance. High gauge factor semiconductor strain-gauges require high temperature microfabrication processes which are incompatible with polymer materials. We therefore developed a MEMS microfabrication platform that separates the high temperature processes from the polymer processes. Figure 2a shows the schematic structure of a self-sensing AFM cantilever made with this hybrid polymer/ceramic/semiconductor fabrication process.

We use a polymer as the main structural component to obtain thick yet soft cantilevers. The strain sensing elements are integrated far away from the neutral axis to maximise the deflection sensitivity. The design incorporates the polymer core enveloped by two hard thin film layers (Figure 2a), optimizing the seamless transmission of strain from the core to the strain sensors[17]. In this trilayer structure, the active electronic parts are embedded between the polymer and the ceramic hard layer, and hence isolated from the environment. This makes the cantilevers inherently fluid compatible and allows for coating of the cantilever's tip side with multifunctional coatings as established for conventional optical beam deflection cantilevers. The fabrication process of the trilayer cantilevers (Figure 2b and supplementary material S3) is based on polymer bonding of two pre-processed wafers, each containing one of the ceramic thin films. The high temperature processes required to fabricate the sensing elements are performed on one or both wafers before wafer bonding. The wafers are then spin-coated and bonded using the polymer Benzocyclobutene (BCB). The release of the devices is achieved by etching through the wafer with KOH and dry etching the trilayer structure. This results in a trilayer cantilever on a silicon chip, where the sensing elements and electrical connections are hermetically sealed inside the hard films (Figure 2b).

The trilayer design provides additional degrees of freedom to optimize the performance of the MEMS cantilever. In traditional single layer cantilevers, only the thickness and planar dimensions can be tuned to obtain a particular MEMS device. In the trilateral devices, the thickness of the BCB core, the thickness of the hard thin film, and the material of the thin film can be tuned to optimize the mechanical and electrical performance of the cantilever. The influence of these three factors can be approximated by a structural mechanics model to calculate the expected deflection sensitivity, spring constant and force sensitivity of the cantilevers (see supplementary material S2). Figure 2c illustrates the theoretical curves representing the deflection sensitivity, spring constant, and force sensitivity of trilayer cantilevers. These cantilevers possess a footprint of 150×50 μm, with a polysilicon strain sensor and two 20 nm low-stress silicon nitride films serving as the hard outer layers. The curves demonstrate the variations in these parameters for different thicknesses of the BCB core. Measurements of two cantilevers fabricated with these parameters, one with 1.6 μm BCB thickness and one with 3.2 μm BCB thickness matched the predicted values very well without any parameter fitting. Using the same model, we compared the theoretical force sensitivity of various versions of trilayer cantilevers with traditional single crystal silicon cantilevers. Figure 2d shows the well-known general trend of increased force sensitivity for decreased thicknesses remains true. However, for a given cantilever thickness, the force sensitivity of the trilayer cantilevers is up to ten times higher than that of silicon cantilevers. In very thin cantilevers, the force sensitivity advantage of the trilayer cantilevers over silicon cantilevers is less pronounced, because the relative stiffness contribution of the polymer decreases compared to the contribution of the silicon nitride.

An inherent advantage of our trilayer process lies in enabling the production of polymer-core cantilevers with strain sensors boasting the same gauge factor found in their silicon counterparts. Notably, trilayer and silicon cantilevers achieve equivalent gauge factors and voltage noise levels by utilizing identical readout electronics. Consequently, the trilayer cantilevers exhibit comparable noise levels while delivering superior force sensitivity when compared to silicon levers. We compared both technologies experimentally by measuring the force noise spectra of two cantilevers with equal dimensions (330 μm long, 110 μm wide and 3.2 μm thick) based on silicon single crystal piezoresistors, both arranged in a Wheatstone bridge configuration (Figure 2e). The trilayer cantilever has a 6 times better force noise compared to the silicon cantilever. The high deflection sensitivity and force sensitivity allow low-noise AFM measurements (Figure 2f) of highly ordered pyrolytic graphite (HOPG) surface, clearly observing the 3.4 Angstrom atomic steps with a Z noise level of 0.4 Angstrom (Figure 2g).

The use of a polymer as the core of the AFM cantilever has, in addition to the increased sensitivity, also benefits for the imaging speed in amplitude modulation (AM) tapping mode. The bandwidth of a cantilever in AM mode is a measure of the maximum rate of topography change the cantilever can accurately detect. The bandwidth scales with $f_0/Q$, where $f_0$ is the cantilever's resonance frequency and Q is its mechanical quality factor (Q-factor)[18]. We previously showed that making cantilevers out of the SU-8 polymer drastically increases the achievable imaging speed[15]. The same holds true for the trilayer cantilevers where the damping is dominated by the polymer core. This is particularly advantageous when imaging in vacuum, where, because of the absence of fluid or air damping, the Q-factor is dominated by the internal damping of the material. We therefore compared the achievable imaging speeds when using the cantilevers in a combined AFM/SEM system (figure 3). We imaged the same sample (wasp eye) with two cantilevers of similar resonance frequency and size (one silicon, one trilayer) using the same AFM (AFSEM, GETec Microscopy GmbH, Austria) installed inside an SEM (figure 3a and b). The SEM reveals the closely packed ommatidium lens surfaces of the wasp eye. The AFM image shows the nano-nipple arrays on the cornea of one ommatidium[19] imaged using a trilayer and a silicon cantilever at 2 lines/sec and 32 lines/sec (Figure 3b). While the silicon cantilever clearly tracks the nano structures poorly at 32 lines/sec scan rate, the trilayer cantilever detects the sample topography significantly better thanks to its lower Q-factor.

A third benefit of our trilayer microfabrication platform is that all sensing elements and electrical connections are hermetically sealed inside the MEMS device, which makes the trilayer devices inherently compatible with measurement applications in fluids. This is particularly important for biological measurements in life science, but also for operating the devices in opaque or harsh chemical environments. As a proof of principle, we imaged the etching process of a polished nickel surface in ferric chloride, a strongly corrosive opaque solution (Figure 4a and supplemental video). Even after 5 hours of imaging, the cantilever showed no signs of degradation. In addition to imaging in liquids, the insulated sensing electronics make the trilayer cantilevers a versatile tool for many other AFM modes: those that require special coatings on the tip, such as Kelvin probe force microscopy (KPFM), or magnetic force microscopy (MFM). Coating traditional self-sensing cantilevers shorts out the self-sensing electrical connections unless additional passivation layers are applied[20,21]. However, such passivation layers negatively affect the self-sensing performance and are prone to failure[22,23]. In the case of trilayer cantilevers, a conductive or magnetic coating can simply be applied through evaporation and sputtering, in the same way as for passive (optical) cantilevers. This enables KPFM and MFM measurements with self-sensing cantilevers. KPFM relies on measuring the potential difference between a conductive tip and the sample surface, which reveals a surface work function map. We have performed frequency modulation self-sensing KPFM on few-layered molybdenum disulfide ($MoS_2$), revealing the sample topography and its surface potential simultaneously (Figure 4b). MFM measurements require a magnetic coating on the AFM cantilever tip. We evaporated 70 nm of $Ni_{81}Fe_{19}$ onto trilayer cantilevers and obtained correlated SEM-AFM-MFM images of interconnected and disconnected networks of $Ni_{81}Fe_{19}$ nanorods patterned on five-fold rotationally symmetric

Penrose P2 quasicrystal lattices (Figure 4c). Such structures, where each nanorod is in essence a small ferromagnet, are candidates for ultra-high-density data storage[24]. The MFM data reveal that the intensity of the magnetic field, displayed in red and blue colours, is different at each of the vertices. The vertices with high intensity act as hotspots where ferromagnetic switching of the nanorods will begin under an applied magnetic field[25]. The permalloy-coated, self-sensing cantilever enabled seamless SEM/AFM/MFM correlative imaging.

The trilayer technology is not limited to self-sensing cantilevers. We fabricated fluid-compatible membrane-type surface-stress sensors (MMS[26], Figure 5a) using the same technology. Such sensors feature a large membrane, suspended by four bridges that contain strain sensors. The membrane can be functionalized to detect different gases or specific molecules. Upon exposure to the target entity, the membrane is subject to a surface stress which is amplified in the suspension bridges and detected with the strain sensors therein. Here, we performed a proof-of-concept experiment where we applied a force at the center of the membrane using an AFM cantilever. Simulations show that for a 2 μN force, a membrane deflection of 50 nm is expected, along with resistive changes of $-4 \times 10^{-4}$ and $2 \times 10^{-4}$ for the parallel and transverse sensors, respectively (Figure 5b). The experimental results confirm these findings (Figure 5c). As trilayer devices are inherently fluid compatible, these membranes could be used for bio-sensing in liquid for point-of-care diagnostics[27,28].

**Discussion**

Integration of self-sensing (and actuation) electronics into MEMS devices is generally performed by depositing the electronic materials on the main structural MEMS material. The advantage of this approach is that a myriad of standard microfabrication processes and materials are available. A conceptual problem for this approach is that the structural material must be able to withstand the often-harsh processing conditions of the electronic materials. This excludes polymers and other more sensitive materials to be used in the structural components of the MEMS device. With the fabrication platform we present in this paper, we overcome this problem by separating the high-temperature processes for the electronic components from the polymer-based processes of the core MEMS material. The trilayer fabrication process has a number of distinct advantages that make it a very promising fabrication platform for advanced MEMS devices. For one, the ability to use polymers as a main structural material extends the Young's modulus and density range for the MEMS body materials by orders of magnitude. This gives additional degrees of freedom to tune the mechanical performance of the MEMS device and complements the traditional geometric optimization degrees of freedom. Second, the electronic elements are no longer on the exposed side of the MEMS device but embedded and hermetically sealed inside the MEMS device. This is particularly beneficial for MEMS devices operating in harsh environments, liquids, or complex biological fluids. Third, the process is inherently expendable allowing for multiple planes of active electronic components inside a MEMS device (5, 7, 9,… etc. layers, each individually electrically addressable). We made use of the above-mentioned process advantages to improve the performance of self-sensing AFM cantilevers as a first group of MEMS devices based on this microfabrication platform, but the technology is not limited to cantilever applications.

The use of polymer materials as the main structural component for self-sensing MEMS can have advantages and disadvantages, depending on the application. The inherently low Q-factor of the polymer-based MEMS devices is advantageous for dynamic AFM applications but is poorly suited for resonators used for mass sensing, where a high Q-factor is important to obtain high sensitivity. The glass transition temperature of BCB (350 °C) limits the temperature range where the MEMS devices can be used. Excessive changes in temperature can change the mechanical properties of the device and for example shift the resonance frequency of the cantilevers. Device aging is also a concern for polymer MEMS. Thus far we have not yet performed systematic aging studies, however we have not observed any excessive aging, even for devices fabricated four years ago.

The opportunities offered by this trilayer fabrication approach extend well beyond the improvement of the sensitivity of self-sensing MEMS devices. In our current work, we only embedded simple piezoresistive strain-gauges into the devices. The platform however would allow integration of more complex electronic devices (for example pre-amplification electronics) directly into the MEMS device, given that all fabrication processes for electronic components occur before polymer bonding and shaping of the MEMS. The fabrication platform also allows integration of actuators as well as sensing electronics (ongoing work). The trilayer fabrication technology holds immense potential for expansion, including the prospect of bonding to wafers featuring advanced pre-processing techniques and potentially encompassing CMOS-based devices. This forward-thinking approach opens up possibilities for seamlessly integrating trilayer structures into cutting-edge technologies, paving the way for enhanced functionality and performance in future applications. This could result in polymer-based MEMS with advanced sensing and signal conditioning electronics. The polymer itself could also be used to add additional functionality to the MEMS devices. BCB can, for example, be etched or photo-patterned[29] before the wafer bonding process. This way, microfluidics could be integrated into the self-sensing MEMS devices, yielding a whole new class of active, microfluidic fluidic MEMS devices.

**Methods**

**1. Cantilever characterization:** To calculate the cantilever properties in figure 2 (c) and (d), we used the following values where E stands for the Young modulus: $E_{LSNT}$=240 GPa, $E_{BCB}$=2.9 GPa, $E_{Silicon}$=130 GPa, and $E_{SiO2}$=66 GPa. Cantilever length=150 μm, width=50 μm and LSNT thickness=20 nm. The BCB thickness changes from 300 nm to 4 μm. The piezoresistors length, width and thickness were 40 μm, 8 μm and 100 nm respectively. The gauge factor of polysilicon was measured at 25.

The experimental data were taken using a Bruker NanoscopeV controller and MultiModeV AFM system. The differential signal from the Wheatstone bridge was amplified first by a low noise instrumentation amplifier (AD8429, Analog Devices, USA), and then two operational amplifiers for a total gain of 1000. The electronics output (deflection signal) was then fed into the Bruker Signal Access Module III. The electrical deflection sensitivity for each individual cantilever was obtained in contact mode. The thermomechanical tuning was measured to characterize the resonance frequency and the spring constant of the cantilevers.

**2. Noise measurement:** Noise spectrum in figure 2 (e) was acquired with a Zurich Instrument, UHF 600MHz, 1.8GSa/S lock in amplifier for a trilayer cantilever and a silicon cantilever (AMG Technology Ltd, Botevgrad, Bulgaria), where both cantilevers have integrated boron doped silicon piezoresistors.

The Amplitude Modulation AM-AFM noise in figure 3(f) was measured with the system described in methods section 1 using the following procedure: the scan size was set to a very small value (e.g. 0.01 nm) and feedback gains were reduced close to zero. This way, there is no topography change and no tracking by the PID. All the fluctuations in the self-sensing deflection signal are contained in the amplitude error signal. The distribution of these fluctuations is used to compute the RMS noise.

**3. Measurements in vacuum:** All the vacuum measurements were performed in an SEM-AFM hybrid system (GETec Austria, moved to QD microscopy, Germany) and Anfatec controller (Anfatec Instrument AG, Germany).

**4. Nickel etch:** The experiment was performed using a Bruker NanoscopeV controller and a Dimension Icon AFM scan head with a homebuilt, liquid-compatible cantilever holder. The electrical deflection signal was sent to the IN0 port of the Bruker Signal Access Module III. The images were taken in PeakForce Tapping with a 50 nN force set point, 1 kHz PeakForce frequency and 1 Hz scan rate.

**5. KPFM:** The image was taken with a Bruker NanoscopeV controller and MultiModeV AFM system in FM-KPFM. A Zurich Instruments UHFLI lock-in amplifier was used to implement the KPFM. The conductive tip of the cantilever was biased with 2.5 V at a frequency of 2 kHz. The cantilever oscillation amplitude at the side-band frequencies was detected and minimized by applying a DC offset voltage to the sample. The control was achieved with a PID controller of the Zurich Instruments UHFLI.

**6. MFM:** Images were taken with the system described in the method section 3.

**7. AFM image processing:** Images were processed in Gwyddion. We removed line-by-line offset using a median correction method and subtracted the background tilt or bow using first and second order polynomial fittings. The nickel etch images were cropped to compensate for the sample drift. Noise in the height images of KPFM and MFM was reduced with a 3-pixel median average filter.

**8. Sample preparation:** The wasp was found dead, and the head was removed and coated with gold and palladium to provide a conductive layer for SEM. The nickel surface was polished with silica suspensions (0.05 μm) in the Interdisciplinary Centre for Electron Microscopy (CIME) at EPFL. The MFM sample was provided by Professor Dirk Grunder (Laboratory of Nanoscale Magnetic Materials and Magnonics, EPFL).

**Acknowledgment**

The authors sincerely thank the Center of Micronanotechnology (CMI) at EPFL for their invaluable assistance throughout the microfabrication process. Additionally, thanks to Professor Andreas Kis and Doctor Hyungoo Ji from the Laboratory of Nanoscale Electronics and Structures (LANES) at EPFL for graciously providing the $MoS_2$ sample for the KPFM measurement.

Furthermore, the authors gratefully acknowledge the financial support received from multiple sources. This includes the European Union FP7/2007-2013/ERC funding under Grant Agreement No. 307338-Eurostars E! 8213-Triple-S, the ERC-2017-CoG, and the InCell project with Project number 773091. Their generous support has been instrumental in completing this research endeavor.

**Contributions**

N. H. and M. N. developed the microfabrication process, fabricated the trilayer MEMS devices in the cleanroom, designed experiments, built instrumentation, performed experiments, analysed data and wrote the paper. J. D. A. developed the microfabrication process and fabricated the trilayer MEMS devices in the cleanroom. S. H. A. and O. P. fabricated the trilayer MEMS devices in the cleanroom.

M. W. performed the SEM/AFM/MFM experiment. M. C G., V. Sh. B., and D. G. provided the $Ni_{81}Fe_{19}$ nanorods MFM sample. G.E.F. coordinated research, designed experiments and wrote the paper.

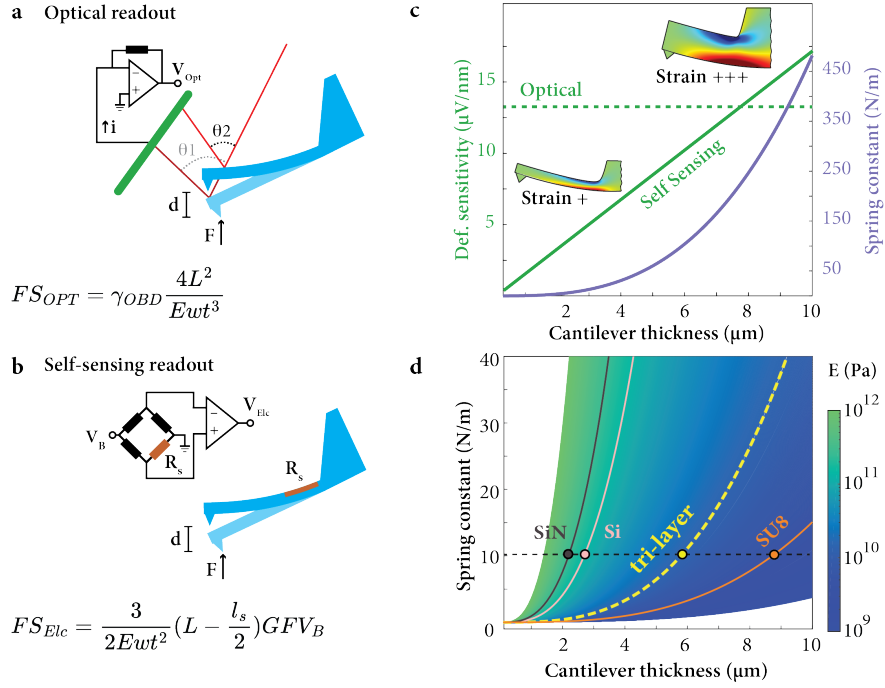

Figure 1 – From force to voltage: transduction and interdependencies. (a) Optical scheme: the applied force causes a deflection of the cantilever, which consequently changes the spot position of the reflected laser beam on the quadrant photodiode. Force sensitivity ($FS_{Opt}$) is defined as $V_{Opt}/F$, where $L$, $w$, $t$ and $E$ are the length, width, thickness and Young modulus of the cantilever, respectively. All parameters that are independent of the cantilever mechanics are combined in one constant $\gamma_{OBD}$ (b) Self-sensing scheme: the applied force/deflection induces strain at the base of the cantilever. A piezoresistive sensor is integrated at the upper surface of the cantilever. The resistance $R_s$ is measured by means of Wheatstone bridge and subsequent read-out electronics. The self-sensing force sensitivity ($FS_{Elc} = V_{Elc}/F$) depends on the gauge factor ($GF$) of the sensing element, the bridge bias voltage ($V_B$), the cantilever dimensions, and the piezoresistor length ($l_s$). (c) The deflection sensitivity of the optical scheme ($V_{Opt}/d$) is independent of the cantilever thickness. The self-sensing deflection sensitivity ($V_{Elc}/F$) increases for larger cantilever thicknesses: a given deflection will induce a higher strain in a thick cantilever, as shown in the two finite element analysis insets. The deflection sensitivity is simulated for a 150 µm×50 µm cantilever footprint and a Wheatstone bridge with 2V bias voltage and 1000 signal amplification. The spring constant of the cantilever however increases with the cube of the thickness. (d) The spring constant also depends on the material's Young modulus. Soft materials like polymers show the same spring constant for higher thicknesses than conventional MEMS materials (e.g., silicon and silicon nitride). The dashed line represents the spring constant of trilayer cantilevers with a footprint size of 150 µm×50 µm.

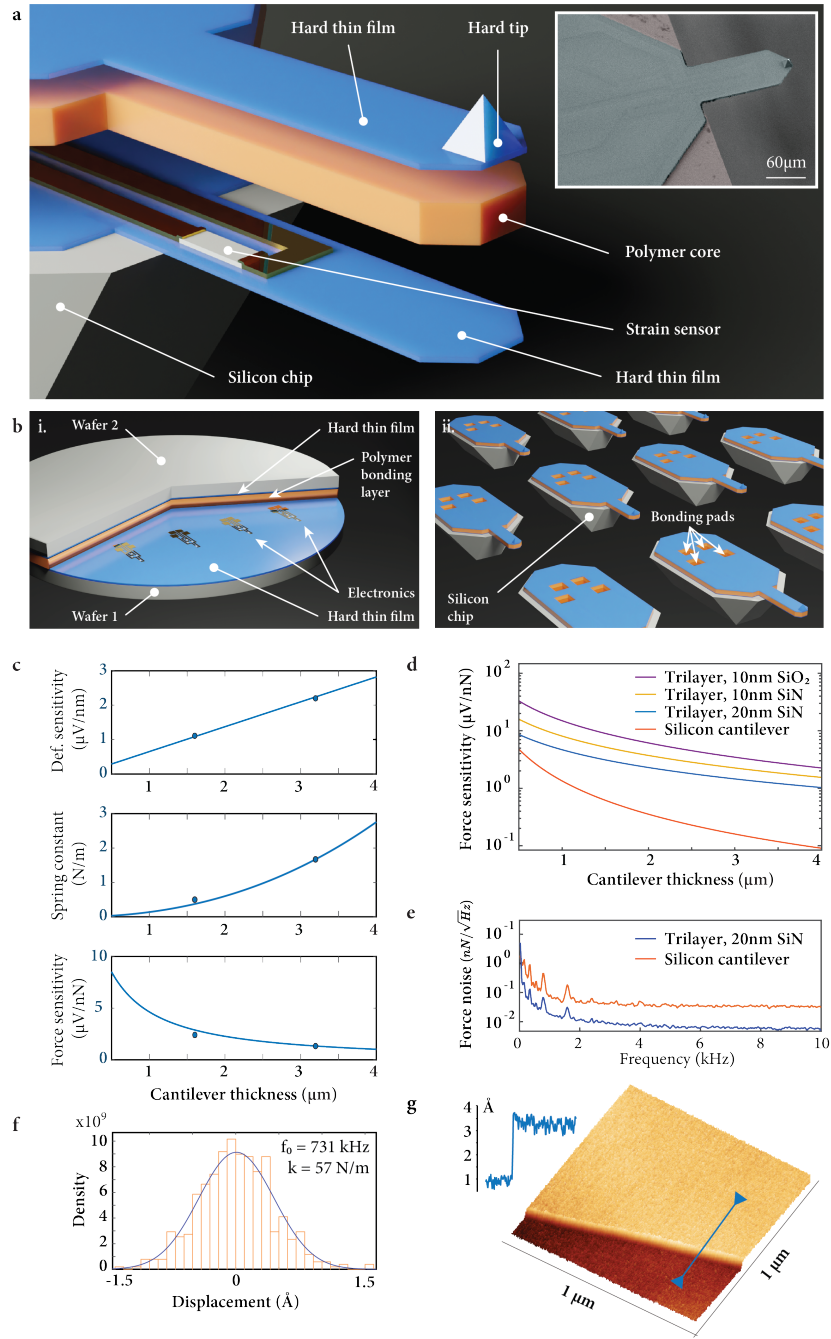

Figure 2 – Trilayer technology concepts and performance. (a) A schematic of the trilayer cantilever illustrates the polymer core and the self-sensing electronics sandwiched between two hard thin films. Thanks to the polymer core, the cantilever can be thick while keeping a low spring constant. The deflection/force sensitivity increases as the sensing element is placed further away from the neutral axis. Inset: SEM image of a trilayer cantilever. The sensing elements are buried under the silicon nitride thin film. (b) The fabrication process is based on polymer bonding of two processed wafers. Each wafer is coated with a thin film of silicon nitride of equal thickness (blue).

BCB (orange) is spin-coated on wafer 1, while piezoresistors and metallic interconnections are patterned on wafer 2. The two wafers are bonded together, and silicon chip bodies (grey) are made through KOH etching. (c) Theoretical and experimental evaluation of the trilayer technology with polysilicon piezoresistors. Circular points represent two trilayer cantilevers with the same planar dimensions (150 μm×50 μm), one with 1.6 μm BCB and the other with 3.2 μm BCB thicknesses. The silicon nitride layer is 20 nm thin. The deflection sensitivity and the spring constant increase by increasing the thickness. The force sensitivity decreases for thicker cantilevers. (d) Comparison between a monolithic silicon cantilever and the trilayer technology with different material combinations. All cantilevers are assumed to have identical strain sensors. The cantilever footprint is 150 μm×50 μm. (e) Experimental proof of better force noise for a trilayer cantilever compared to a silicon cantilever with similar dimensions and piezoresistors (e.g., monocrystalline silicon). (f) Thanks to its enhanced deflection sensitivity, an RMS noise value of 0.4 Å was obtained for a trilayer cantilever with integrated polysilicon piezoresistors in the air with k = 57 N/m. (g) A single atomic layer of HOPG is resolved by the trilayer cantilever.

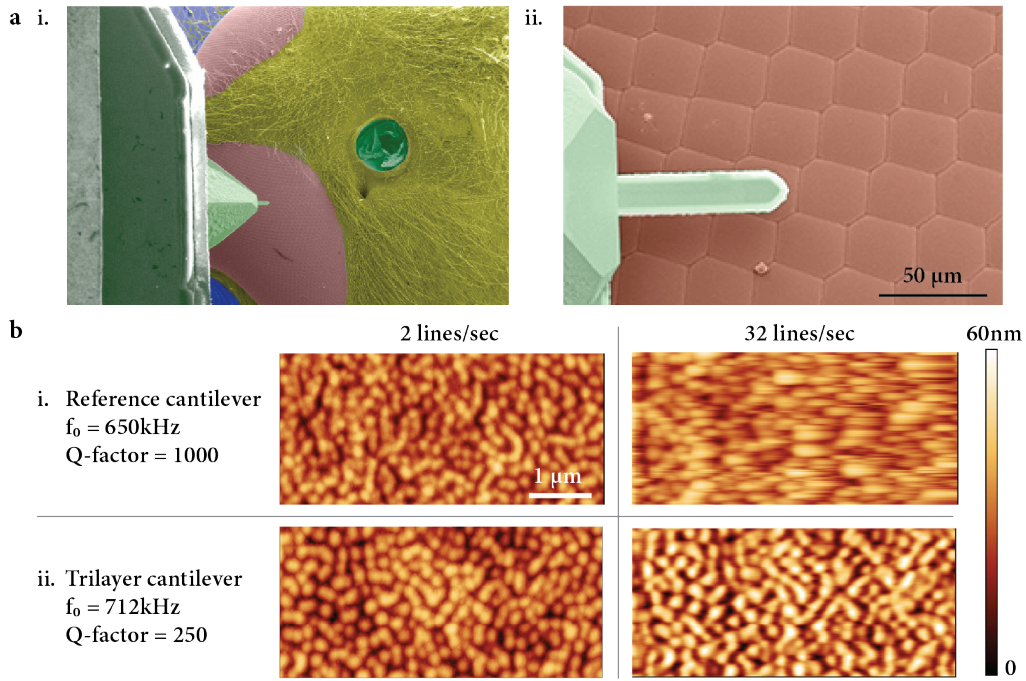

Figure 3 – High tracking bandwidth of trilayer cantilevers for AM-AFM in vacuum. (a) i - A wasp eye is investigated with an SEM-AFM hybrid system. ii - SEM image of the cantilever and the closely packed ommatidia at the surface of the wasp eye. The SEM is used to navigate the cantilever on top of an ommatidium. (b) The ommatidium surface was imaged with two different cantilevers. The trilayer cantilever has a lower Q-factor and therefore higher detection bandwidth than the reference silicon cantilever. It shows significantly better tracking at an imaging speed of 32 lines per second.

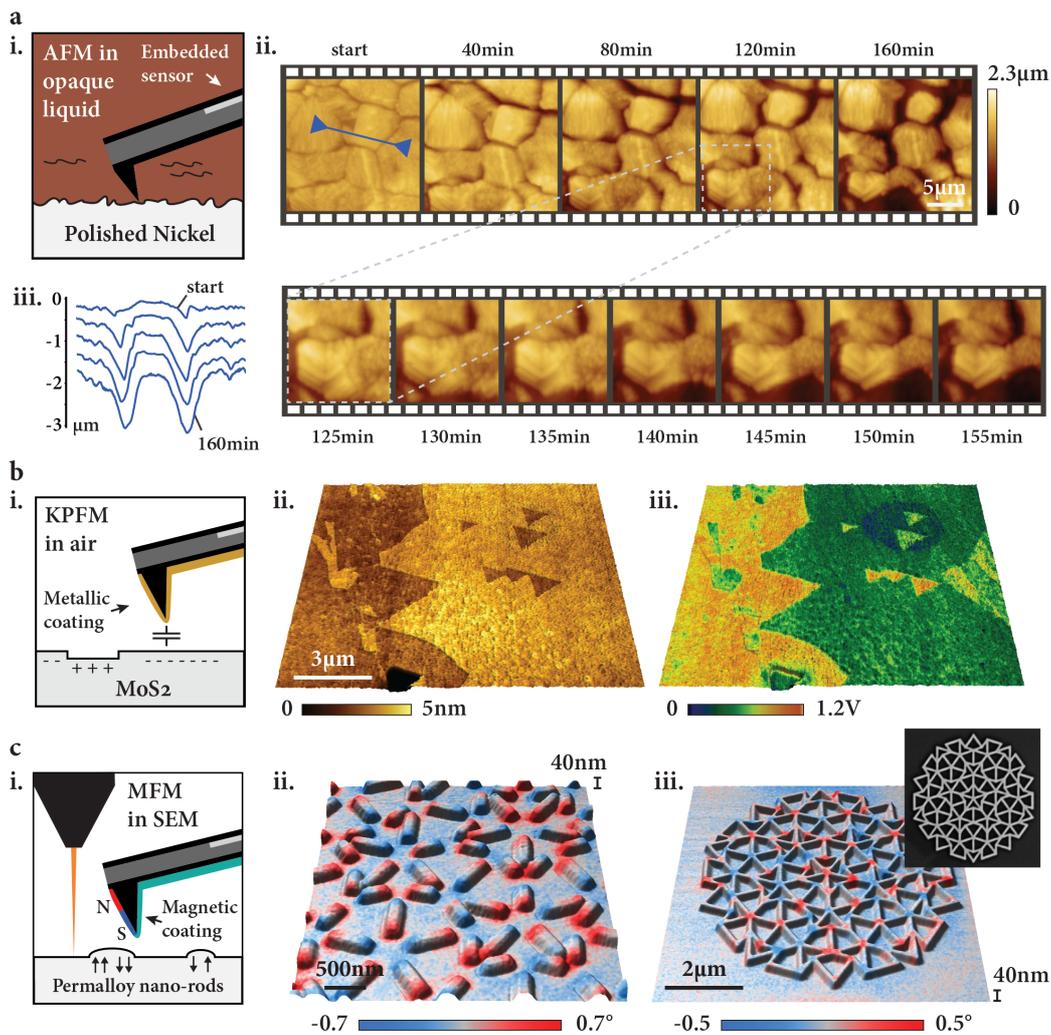

Figure 4 – Trilayer cantilevers as a versatile platform for different SPM techniques. (a) i - A trilayer cantilever is immersed in FeCl$_3$ to investigate how a polished nickel surface evolves when it is exposed to the corrosive fluid. ii – A 160-minute time-lapse shows how the nickel grains are etched by FeCl$_3$. iii – A line profile reveals how grain boundaries evolve during etching. (b) i - A trilayer cantilever is coated with 100 nm of gold to provide a conductive tip for KPFM applications. ii – Sample topography of few-layered MoS$_2$ showing two distinct layers. iii - Superposition of topography and work function reveals the surface potential difference between the layers. (c) i - A trilayer cantilever is coated with 70 nm of Ni$_{81}$Fe$_{19}$ to perform MFM in vacuum with an SEM-AFM hybrid system. ii - Superposition of topography and phase data reveals the intensity of the magnetic field created by separated Ni$_{81}$Fe$_{19}$ nanorods. iii – The same technique is applied to interconnected nanorods. The inset shows an SEM image of the nanorod structure.

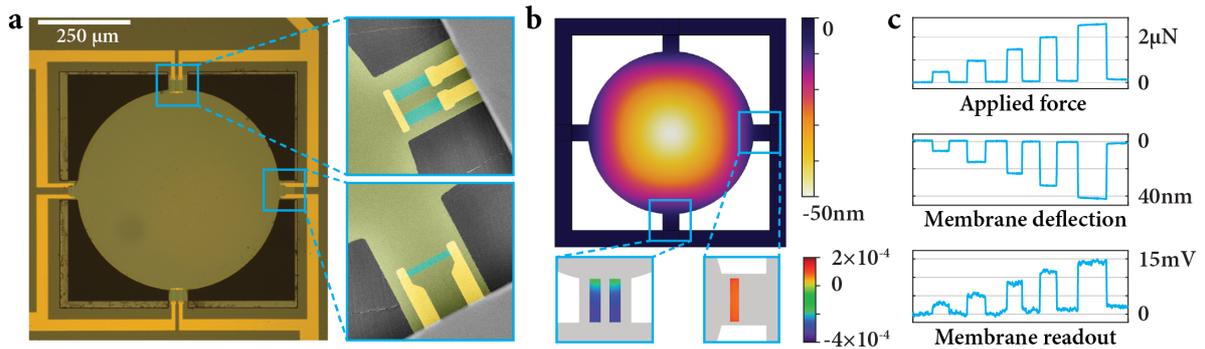

Figure 5 – Trilayer MEMS for fluid-proof membrane surface stress sensing. (a) A trilayer membrane with a diameter of 500 µm is suspended by four beams with integrated piezoresistive sensors. In two beams, the resistors are parallel, while they are transverse in the other two. (b) In a proof-of-concept experiment, a point force is applied at the center of the membrane. The finite element analysis shows negative resistive change in the parallel piezoresistors and positive change in the transverse resistors. (c) The force was applied in the middle of the membrane using an AFM cantilever, inducing a deflection of the membrane. The resistive change of the piezoresistors is detected with a full Wheatstone bridge readout.